\documentstyle[pra,aps,floats]{revtex}

\newcommand{\smallfrac}[2]{\frac{\mbox{\scriptsize $#1$}}{\mbox{\scriptsize $#2$}}}
\newcommand{\bfe}{{\bf e}}
\newcommand{\bfq}{{\bf q}}
\newcommand{\bfs}{{\bf s}}
\newcommand{\bft}{{\bf t}}
\newcommand{\bfw}{{\bf w}}
\newcommand{\bfx}{{\bf x}}
\newcommand{\bfy}{{\bf y}}
\newcommand{\bfz}{{\bf z}}
\newcommand{\bfsigma}{{\mbox{\boldmath $\sigma$}}}
\newcommand{\bfnabla}{{\mbox{\boldmath $\nabla$}}}
\newcommand{\caln}{{\cal N}}
\newcommand{\calo}{{\cal O}}
\newcommand{\sgn}{{\mbox{sgn}}}
\newcommand{\git}[4]{\int d#2\; T(#1\rightarrow #2)\sqrt{\frac{f(#2)#3}{f(#1)#4}}}
\newcommand{\fl}{f_{\mbox{\tiny $L$}}}
\newcommand{\fh}{f_{\mbox{\tiny $H$}}}

\begin{document}

\title{
  \begin{flushleft}
    {\footnotesize BU-CCS-980301}\\
    {\footnotesize To appear in {\it Physical Review E}}\\[0.3cm]
  \end{flushleft}
  \bf A Generalization of Metropolis and Heat-Bath Sampling for Monte Carlo Simulations}
\author{
  Bruce M. Boghosian\\
  {\footnotesize Center for Computational Science, Boston University,}\\
  {\footnotesize 3 Cummington Street, Boston, Massachusetts 02215, U.S.A.}\\
  {\footnotesize{\tt bruceb@bu.edu}}\\[0.3cm]
  }
\date{\today}
\maketitle

\section*{Abstract}
For a wide class of applications of the Monte Carlo method, we describe
a general sampling methodology that is guaranteed to converge to a
specified equilibrium distribution function.  The method is distinct
from that of Metropolis in that it is sometimes possible to arrange for
unconditional acceptance of trial moves.  It involves sampling states in
a local region of phase space with probability equal to, in the first
approximation, the square root of the desired global probability density
function.  The validity of this choice is derived from the
Chapman-Kolmogorov equation, and the utility of the method is
illustrated by a prototypical numerical experiment.

\vspace{0.2truein}

\par\noindent {\bf Keywords}: Monte Carlo, Metropolis sampling, detailed
balance, thermodynamic equilibrium

\vspace{0.2truein}

In Monte Carlo calculations, it is often necessary to sample from a
probability density function (PDF) that is known only to within a
multiplicative constant.  That is, we often know an analytic form for
\[
F(\bfx) = Z f(\bfx),
\]
where $f(\bfx)$ is the PDF from which we would like to sample, and where
the constant $Z$ is unknown.  In computational statistical physics, for
example, it is often necessary to sample from the Boltzmann-Gibbs
distribution
\begin{equation}
  f(\bfx) = \frac{1}{Z}\exp\left[-\beta H(\bfx)\right],
  \label{eq:bg}
\end{equation}
where $\bfx$ coordinatizes the phase space, $H(\bfx)$ is the system
Hamiltonian, and $\beta$ is the inverse temperature.  A simple analytic
form is generally available for $H(\bfx)$ but not for $Z$.  Indeed, if
such a form were known for $Z$, it would not be necessary to resort to
Monte Carlo methods.

One way to break this apparent impasse was provided by an ingenious
algorithm due to Metropolis {\it et al.}~\cite{bib:mmc} in 1953.  This
algorithm provides a Markov process on the state space whose
time-asymptotic state is the desired PDF, $f(\bfx)$, but whose
transition kernel $K(\bfy\rightarrow\bfx)$ involves only {\it ratios} of
this PDF.  It does this by noting that a sufficient condition for a
stationary solution of the Chapman-Kolmogorov equation for the Markov
process,
\begin{equation}
  f(\bfx) = \! \int \! d\bfy\, f(\bfy) K(\bfy\rightarrow\bfx),
  \label{eq:ck}
\end{equation}
is that of {\it detailed balance},
\begin{equation}
  f(\bfx)K(\bfx\rightarrow\bfy) = f(\bfy)K(\bfy\rightarrow\bfx).
  \label{eq:db}
\end{equation}
To see this, we need only integrate both sides of Eq.~(\ref{eq:db}) with
respect to $\bfy$, using the normalization constraint on the transition
kernel,
\begin{equation}
  \int \! d\bfy\, K(\bfx\rightarrow\bfy) = 1;
  \label{eq:knorm}
\end{equation}
the result is Eq.~(\ref{eq:ck}).  We then note that the choice
\begin{equation}
  K(\bfy\rightarrow\bfx) = T(\bfy\rightarrow\bfx)
  \min\left[1,
    \frac{f(\bfx) T(\bfx\rightarrow\bfy)}{f(\bfy) T(\bfy\rightarrow\bfx)}
  \right]
  \label{eq:tpmet}
\end{equation}
satisfies Eq.~(\ref{eq:db}) and involves only ratios of the PDF.

A simple application of the Metropolis method to a statistical physics
problem might then begin with state $\bfy$ and select a new state $\bfx$
according to some {\it trial PDF}, $T(\bfy\rightarrow\bfx)$.  A convenient
(but not necessary) condition on the trial PDF is
\begin{equation}
  T(\bfy\rightarrow\bfx) = T(\bfx\rightarrow\bfy).
  \label{eq:sym}
\end{equation}
This new trial state is then {\it accepted} with probability
\[
\min\left\{1,\exp\left[-\beta\left(H(\bfx)-H(\bfy)\right)\right]\right\},
\]
so that the net transition probability is given by Eq.~(\ref{eq:tpmet}).
Thus, the trial state is accepted with probability unity if it results
in a decrease of energy, and with probability $\exp\left(-\beta\Delta
  H\right)$ if it results in an energy increase of $\Delta H$. If the
trial is not accepted, then the current state is retained.  More
sophisticated implementations of the Metropolis algorithm bias the
selection of a trial state so that Eq.~(\ref{eq:sym}) is
violated~\cite{bib:frenkel}, but the detailed-balance condition,
Eq.~(\ref{eq:db}), is maintained.

In this paper, we describe a fundamentally different kind of Markov
process for the sampling of a PDF $f(\bfx)$ that is known only to within
a multiplicative constant.  The method is distinct from that of
Metropolis in that moves may, in some circumstances, be accepted
unconditionally.  Nevertheless, it admits $f(\bfx)$ as a time-asymptotic
solution of the Chapman-Kolmogorov equation, Eq.~(\ref{eq:ck}).  The
proposed Markov process uses a transition rate of the form
\begin{equation}
  K(\bfy\rightarrow\bfx) =
  \frac{T(\bfy\rightarrow\bfx) g(\bfx)}
  {\int \! d\bfz\, T(\bfy\rightarrow\bfz) g(\bfz)},
  \label{eq:tp}
\end{equation}
where $T(\bfy\rightarrow\bfx)$ is the trial PDF, and $g(\bfx)$ is to be
determined in terms of the desired time-asymptotic state $f(\bfx)$.  Note
that this form obeys Eq.~(\ref{eq:knorm}) manifestly.  Essentially, it tells
us to sample from $g(\bfx)$, normalized over only those states accessible by
the trial PDF $T(\bfy\rightarrow\bfx)$.  We note that the dependence of
$g(\bfx)$ on $f(\bfx)$ may well be complicated (nonlinear and/or nonlocal),
but it must be such that the transition kernel of Eq.~(\ref{eq:tp}) is
unchanged when $f(\bfx)$ is scaled by a multiplicative constant.

If we insist that the trial process be symmetric in the sense of
Eq.~(\ref{eq:sym}), then it is straightforward to verify that the
Chapman-Kolmogorov equation, Eq.~(\ref{eq:ck}), with transition kernel given
by Eq.~(\ref{eq:tp}), is satisfied by
\begin{equation}
  f(\bfx) = g(\bfx)\! \int \! d\bfy\, T(\bfy\rightarrow\bfx) g(\bfy).
  \label{eq:inteq}
\end{equation}
Thus we must invert Eq.~(\ref{eq:inteq}) to get $g(\bfx)$ in terms of
$f(\bfx)$.  Eq.~(\ref{eq:inteq}) is a nonlinear integral equation for
$g(\bfx)$, but the reward for inverting it is a Markov process with specified
trial PDF, $T(\bfy\rightarrow\bfx)$, and stationary state $f(\bfx)$.  The
utility of Eq.~(\ref{eq:inteq}) is the central observation of this paper.

One limit in which Eq.~(\ref{eq:inteq}) has a trivial solution is that
for which all states are equally accessible.  If
$T(\bfy\rightarrow\bfx)$ is independent of $\bfx$, it is easy to see
that Eq.~(\ref{eq:inteq}) is solved by $g(\bfx)\propto f(\bfx)$.  This
solution is the {\it heat-bath} Monte Carlo algorithm.  The proposed
method can thus be regarded as a generalization of the heat-bath
algorithm for nonuniform trial PDF's.

We can better understand the character of the solutions to
Eq.~(\ref{eq:inteq}) if we rewrite it as follows:
\begin{equation}
  g(\bfx) =
  \sqrt{
    \frac{f(\bfx)}
    {\int \! d\bfy\,
      T(\bfx\rightarrow\bfy)
      \frac{g(\bfy)}{g(\bfx)}}}.
  \label{eq:ggf}
\end{equation}
Let us suppose that $g(\bfy)$ is reasonably constant over the set of
$\bfy$ for which $T(\bfx\rightarrow\bfy)$ is appreciable.  For
simplicity, let us also suppose that the trial PDF is normalized,
\[
\int \! d\bfy\, T(\bfy\rightarrow\bfx) = 1.
\]
Then the integral in the denominator is nearly unity, and
$g(\bfx)\approx\sqrt{f(\bfx)}$ to first approximation.  In this case, it
makes sense to solve Eq.~(\ref{eq:ggf}) by the following successive
approximation scheme:
\begin{eqnarray*}
  g(\bfx) &=& \lim_{\ell\rightarrow\infty} g^{(\ell)} (\bfx)\\
  g^{(0)}(\bfx) &=& \mbox{constant}\\
  g^{(\ell)} (\bfx) &=&
  \sqrt{
    \frac{f(\bfx)}
    {\int \! d\bfy\,
      T(\bfx\rightarrow\bfy)
      \frac{g^{(\ell-1)}(\bfy)}{g^{(\ell-1)}(\bfx)}}}.
\end{eqnarray*}
The first few approximations to $g(\bfx)$ are then
\begin{eqnarray}
  g^{(1)}(\bfx) &=&
  \sqrt{f(\bfx)}\nonumber\\
  g^{(2)}(\bfx)
  &=&
  \sqrt{
    \frac{f(\bfx)}
    {\git{\bfx}{\bfy}{}{}}}\nonumber\\
  g^{(3)}(\bfx)
  &=&
  \sqrt{
    \frac{f(\bfx)}
    {\git{\bfx}{\bfy}
      {\git{\bfx}{\bfz}{}{}}
      {\git{\bfy}{\bfz}{}{}}}}\nonumber\\
  g^{(4)}(\bfx)
  &=&
  \sqrt{
    \frac{f(\bfx)}
    {\git{\bfx}{\bfy}
      {\git{\bfx}{\bfz}
        {\git{\bfx}{\bfw}{}{}}
        {\git{\bfz}{\bfw}{}{}}}
      {\git{\bfy}{\bfz}
        {\git{\bfy}{\bfw}{}{}}
        {\git{\bfz}{\bfw}{}{}}}}}.
  \label{eq:iter}
\end{eqnarray}
In all cases, note that if $f(\bfx)$ is multiplied by a factor $C$, the
$g^{(\ell)} (\bfx)$ scale by an overall factor $\sqrt{C}$, but this
factor cancels when inserted into Eq.~(\ref{eq:tp}) for the transition
kernel.  Thus the transition kernel depends only on ratios of the PDF,
as promised.

At this point, we make two observations: First, all of the above
considerations apply to a discrete state space, if we just replace the
integrals by sums as appropriate.  Second, all of the above
considerations can be applied to situations for which
$T(\bfx\rightarrow\bfx)=0$, so that the walker steps to another state
with probability unity, though there are no guarantees that
Eq.~(\ref{eq:inteq}) will admit a solution in that case.

As a first simple example of this methodology for which the integral
equation, Eq.~(\ref{eq:inteq}), may be solved {\it without any
  approximation at all}, consider a five-state system with
$\bfx,\bfy\in\{1,2,3,4,5\}$, so that $T(\bfx\rightarrow\bfy)$ may be
represented as a matrix which we take to be
\[
T = 
\left(
  \begin{array}{ccccc}
    0 & \smallfrac{1}{2} & 0 & 0 & \smallfrac{1}{2}\\
    \smallfrac{1}{2} & 0 & \smallfrac{1}{2} & 0 & 0\\
    0 & \smallfrac{1}{2} & 0 & \smallfrac{1}{2} & 0\\
    0 & 0 & \smallfrac{1}{2} & 0 & \smallfrac{1}{2}\\
    \smallfrac{1}{2} & 0 & 0 & \smallfrac{1}{2} & 0
  \end{array}
\right).
\]
Note that the diagonal elements $T(\bfx\rightarrow\bfx)=0$ so that moves
that leave the state unchanged are expressly forbidden.  The analog of
the integral equation, Eq.~(\ref{eq:inteq}), is the set of nonlinear
algebraic equations
\begin{displaymath}
  f_j = g_j (g_{j-1} + g_{j+1}),
\end{displaymath}
where we have used subscripts modulo 5 to denote functional arguments for
simplicity.  We note that scaling the $g_j$ by $C$ and the $f_j$ by $C^2$
leaves the system unchanged, and we search for solutions only to within an
arbitrary multiplicative constant.  It may be verified that the following is
such a solution:
{\scriptsize
\begin{equation}
g_j = C
(+ f_j + f_{j+1} - f_{j+2} + f_{j+3} - f_{j+4})
(+ f_j - f_{j+1} + f_{j+2} + f_{j+3} - f_{j+4})
(+ f_j - f_{j+1} + f_{j+2} - f_{j+3} + f_{j+4}),
\label{eq:analyt}
\end{equation}}
where again all subscripts are taken modulo 5.  The constant $C$ turns
out to be equal to
{\tiny
  \[
  C = \frac{1}{\sqrt{2
      (- f_1 + f_2 - f_3 + f_4 + f_5)
      (- f_1 + f_2 + f_3 - f_4 + f_5)
      (+ f_1 + f_2 - f_3 + f_4 - f_5)
      (+ f_1 - f_2 + f_3 + f_4 - f_5)
      (+ f_1 - f_2 + f_3 - f_4 + f_5)}},
  \]}
but it is unnecessary to know this in order to use this solution, since
only the ratios of the $g_j$'s matter.

To be concrete, suppose that we take $f_1=1$, $f_2=2$, $f_3=3$, $f_4=2$,
and $f_5=1$; obviously, these should be normalized by the factor $1/9$,
but let us pretend that we do not know this normalization factor.  The
above equations give $g_1=g_2=g_4=g_5=3C$, and $g_3=9C$.  Thus, if we
simulate a Markov process with transition matrix
\[
K =
\left(
  \begin{array}{ccccc}
    0 & \frac{g_2}{g_2+g_5} & 0 & 0 & \frac{g_5}{g_2+g_5}\\
    \frac{g_1}{g_1+g_3} & 0 & \frac{g_3}{g_1+g_3} & 0 & 0\\
    0 & \frac{g_2}{g_2+g_4} & 0 & \frac{g_4}{g_2+g_4} & 0\\
    0 & 0 & \frac{g_3}{g_3+g_5} & 0 & \frac{g_5}{g_3+g_5}\\
    \frac{g_1}{g_4+g_1} & 0 & 0 & \frac{g_4}{g_4+g_1} & 0
  \end{array}
\right) =
\left(
  \begin{array}{ccccc}
    0 & \smallfrac{1}{2} & 0 & 0 & \smallfrac{1}{2}\\
    \smallfrac{1}{4} & 0 & \smallfrac{3}{4} & 0 & 0\\
    0 & \smallfrac{1}{2} & 0 & \smallfrac{1}{2} & 0\\
    0 & 0 & \smallfrac{3}{4} & 0 & \smallfrac{1}{4}\\
    \smallfrac{1}{2} & 0 & 0 & \smallfrac{1}{2} & 0
  \end{array}
\right),
\]
the result will be the desired equilibrium state.  Note that this Markov
process is nothing like the Metropolis algorithm, in that trial states
are accepted unconditionally.  Also note that it is distinct from the
{\it heat-bath} algorithm, wherein {\it all} states must sampled with
probability proportional to $f(\bfx)$.

It is of course not always possible to do what we have done in the
simple example above.  If we have one state that is much more probable
than any other, our intuition tells us that a Markov process cannot
force us to leave that state with probability unity every time we are in
it, since the result would be our spending half of our time in a much
less probable state.  To see how the solution for the five-state system
breaks down in that case, let us suppose that state $3$ has a high
probability $\fh$, and the other four states have a low probability
$\fl$.  We find that $g_1 = g_5 = (2\fl-\fh) \fh^2$, $g_2 = g_4 =
(2\fl-\fh)^2 \fh$, and $g_3 = \fh^3$.  We note that
$\sgn(g_1)=\sgn(g_5)$ and $\sgn(g_2)=\sgn(g_3)=\sgn(g_4)$, but that
these signs will not be the same if $\fh>2\fl$.  If they are not the
same, our transition matrix will involve negative probabilities.  Hence
we arrive at another requirement on the solutions of
Eq.~(\ref{eq:inteq}): Any legitimate solution for $g(\bfx)$ must be of
definite sign.

It is one thing to treat a five-state system, but quite another to treat,
say, a two-dimensional Ising model on a $10\times 10$ lattice, for which
there are $2^{100}$ distinct states.  In such a situation, it will be no more
possible to store $g(\bfx)$ on a computer than it is to store $f(\bfx)$.
Thus, we must search for ways to obtain partial solutions to
Eq.~(\ref{eq:inteq}) for large systems {\it on the fly}; more specifically,
if we are in a state $\bfx$, we must be able to easily compute $g(\bfy)$ for
all the states for which $T(\bfx\rightarrow\bfy)$ is nonzero.  This is enough
information to sample the next state and then we can repeat the process.  So
for the remainder of this paper we turn our attention to larger systems.

It is likely that analytic solutions in the spirit of
Eqs.~(\ref{eq:analyt}) exist for arbitrarily large systems, but they
seem to become correspondingly more difficult to find.  What is
interesting is that the iterative process given in Eqs.~(\ref{eq:iter})
actually becomes {\it easier} to solve for larger systems, as long as
the distribution $f(\bfx)$ is smooth in the state space.  To see this,
let us consider an $N$-state system with periodic, nearest-neighbor
state connectivity and
\[
f_j = 1+2\sin^2\left(\frac{\pi j}{N-1}\right)
\]
for $j=0,\ldots,N-1$; note that when $N=5$ this reduces to the numerical
example used above for which we know that the exact solution for the
$g_j$ obeys $g_2/g_1=g_4/g_1=g_5/g_1=1$ and $g_3/g_1=3$.  (Recall that
only {\it ratios} of the $g_j$'s matter.)  These results are not
particularly close to the first approximation of the iterative process
of Eq.~(\ref{eq:iter}) which would be $g^{(1)}_5/g^{(1)}_1 =
\sqrt{f_5/f_1} = 1$, $g^{(1)}_{2,4}/g^{(1)}_1 = \sqrt{f_{2,4}/f_1} =
\sqrt{2}$, and $g^{(1)}_3/g^{(1)}_1 = \sqrt{f_3/f_1} = \sqrt{3}$.  This
is because that iterative process was predicated on the smoothness of
$f(\bfx)$, and there is no obvious sense in which a function defined on
five discrete points is smooth.  Nevertheless, as Table~\ref{tab:tab1}
illustrates, the iteration does appear to converge.
Table~\ref{tab:tab2} shows that when we let $N=50$ (without making the
function $f$ any more rapidly varying), the iteration converges much
faster.  (This is especially evident if we compare the values of the
iteration number $\ell$ in the two tables.)  Moreover, we see that the
asymptotic result is not very different from the first approximation,
which is proportional to the square root of the distribution function.
This suggests that we develop a perturbation theory for $g(\bfx)$ about
$\sqrt{f(\bfx)}$, and we turn our attention to this approach below.
\begin{table}
  \begin{tabular}{l||c|c|c|c|c}
    $g^{(\ell)}_j/g^{(\ell)}_1$ &
    $\ell=1$ & $\ell=25$ & $\ell=50$ & $\ell=75$ & $\ell=100$ \\
    \hline
    \hline
    $j=2,4$ & 1.41421 & 1.20022 & 1.05966 & 1.01876 & 1.00600\\
    \hline
    $j=3$   & 1.73205 & 2.72733 & 2.90754 & 2.96972 & 2.99019
  \end{tabular}
\caption{Iterative solution with $N=5$}
\label{tab:tab1}
\end{table}
\begin{table}
  \begin{tabular}{l||c|c|c|c|c}
    $g^{(\ell)}_j/g^{(\ell)}_1$ &
    $\ell=1$ & $\ell=2$ & $\ell=3$ & $\ell=4$ & $\ell=100$ \\
    \hline
    \hline
    $j=10$ & 1.26302 & 1.26408 & 1.26378 & 1.26361 & 1.26261\\
    \hline
    $j=20$ & 1.66176 & 1.66450 & 1.66410 & 1.66388 & 1.66301\\
    \hline
    $j=30$ & 1.68466 & 1.68747 & 1.68707 & 1.68685 & 1.68598\\
    \hline
    $j=40$ & 1.30976 & 1.31107 & 1.31076 & 1.31059 & 1.31018
  \end{tabular}
\caption{Iterative solution with $N=50$}
\label{tab:tab2}
\end{table}

Let us restrict our attention to continuous state spaces, and suppose
that $T(\bfx\rightarrow\bfy)=W(\|\bfx-\bfy\|)$, for some norm
$\|\cdot\|$, where the region of support of $W(\|\bfy\|)$ is very small
compared to the characteristic scale(s) of $f(\bfx)$.  Then we can
Taylor expand $g(\bfy)$ about the point $\bfx$ in Eq.~(\ref{eq:inteq}).
In particular, if we then suppose that the moments of the function $W$
are
\begin{eqnarray}
  & &\int\! d\bfy\, W(\|\bfy\|) = 1\nonumber\\
  & &\int\! d\bfy\, W(\|\bfy\|)\bfy = \bft\nonumber\\
  & &\int\! d\bfy\, W(\|\bfy\|)\bfy\bfy = \bft\bft + \bfsigma\label{eq:moms}
\end{eqnarray}
where $\bft$ is the mean and $\bfsigma$ is the variance tensor, then
Eq.~(\ref{eq:inteq}) becomes
\begin{eqnarray*}
  f(\bfx) &=& g(\bfx)
  \left[
    g(\bfx) + 
    \bft\cdot\bfnabla g(\bfx) +
    \frac{1}{2}\left(\bft\bft+\bfsigma\right):
    \bfnabla\bfnabla g(\bfx) + \cdots
  \right]\\
  &=&
  \left[
    g(\bfx)
  \right]^2
  \left[
    1 +
    \frac{\bft\cdot\bfnabla g(\bfx)}{g(\bfx)} +
    \frac{\left(\bft\cdot\bfnabla\right)^2 g(\bfx)}{2g(\bfx)} +
    \frac{\bfsigma :\bfnabla\bfnabla g(\bfx)}{2g(\bfx)} + \cdots
  \right].
\end{eqnarray*}
We can invert the above equation perturbatively to find
\begin{eqnarray*}
g
&=&
\sqrt{f}
\left[
  1 - \frac{\bft\cdot\bfnabla f}{4f} +
  \frac{\left(\bft\cdot\bfnabla f\right)^2}{32f^2} -
  \frac{\bfsigma :\bfnabla\bfnabla f}{8f} +
  \frac{\bfsigma :\left(\bfnabla f\right)\left(\bfnabla f\right)}{16f^2} +
  \cdots
\right],
\end{eqnarray*}
where all functions and derivatives are understood to be evaluated at
$\bfx$.  Once again, it is manifest that the transition kernel,
Eq.~(\ref{eq:tp}), computed from this $g(\bfx)$ will depend only on
ratios of $f(\bfx)$.  When the desired distribution function $f(\bfx)$
is of Boltzmann-Gibbs form, as in Eq.~(\ref{eq:bg}), the above equation
reduces to
\begin{equation}
g =
e^{-\beta H/2}
\left[
1 + \frac{\beta}{4}\bft\cdot\bfnabla H +
\frac{\beta^2}{32}\left(\bft\cdot\bfnabla H\right)^2 +
\frac{\beta}{8}\bfsigma :\bfnabla\bfnabla H -
\frac{\beta^2}{16}\bfsigma :\left(\bfnabla H\right)\left(\bfnabla H\right) +
\cdots
\right].
\label{eq:gmol}
\end{equation}
This form is potentially useful for Monte Carlo simulations of molecular
systems, for which the function $H(\bfx)$ is known in analytic form and
the state space is continuous, so that gradients of $H(\bfx)$ can also
be computed in analytic form.  It is interesting that the first
approximation to $g(\bfx)$ in this case is $\sqrt{f(\bfx)}$ which is
also of Boltzmann-Gibbs form, but at {\it twice} the desired
temperature.

To apply Eq.~(\ref{eq:gmol}) to a molecular system with potential energy
\[
V(\bfq) = \frac{1}{2}\sum_{i,j}^N v(|\bfq_i-\bfq_j|)
\]
and PDF $\exp[-\beta V(\bfq)]/Z$, we calculate $g$ from
Eq.~(\ref{eq:gmol}) (with $H\rightarrow V$), and consider moves that
take the walker at $\bfq=(\bfq_1,\ldots,\bfq_N)$ to one at
\[
\bfq+\Delta\bfq=
(\bfq_1+\sigma\hat{\bfe}_1,\ldots,\bfq_N+\sigma\hat{\bfe}_N),
\]
where the $\hat{\bfe}_i$ are spatial unit vectors, sampled with
probability $g(\bfq+\Delta\bfq)$.  Note that all of the particles must
move a distance $\sigma$ in this scheme.  The sampling of unit vectors
$\hat{\bfe}_i$ may be carried out by the Metropolis or any other
suitable method, but after that is completed the new walker position is
$\bfq+\Delta\bfq$ with probability unity.  For sufficiently small
$\sigma$, this method is guaranteed to converge and unconditionally
accepts moves.  A potential pitfall is that this sufficiently small value of
$\sigma$ may be very small indeed.

As a final example, we turn our attention to the Ising model with $N$
spins $\bfs$ and Hamiltonian
\[
H(\bfs) = \frac{1}{2}\sum_i^N\sum_{j\in \caln_i} J s_i s_j,
\]
where $\caln_i$ denotes the neighborhood of spin $i$.  Let us introduce
the operator $\hat{\theta}_i$ such that $\hat{\theta}_i\bfs$ is the
state obtained from state $\bfs$ by flipping spin $i$.  We also denote
the resulting change in observable $F(\bfs)$ by
\[
\hat{\delta}_i F(\bfs) \equiv F(\hat{\theta}_i\bfs)-F(\bfs).
\]
Let us suppose that our trial PDF can take us from state $\bfs$ to any
one of the $N$ states $\hat{\theta}_i\bfs$ for $i=1,\ldots,N$.  Once
again, this condition expressly forbids the system from remaining in its
present state.  The analog of our integral equation,
Eq.~(\ref{eq:inteq}), is then
\[
e^{-\beta H(\bfs)}
=
g(\bfs) \frac{1}{N} \sum_i^N g(\hat{\theta}_i\bfs)
=
g(\bfs) \frac{1}{N} \sum_i^N \left[g(\bfs) + \hat{\delta}_i g(\bfs)\right]
=
\left[g(\bfs)\right]^2
\left[1 + \frac{1}{N} \sum_i^N
      \frac{\hat{\delta}_i g(\bfs)}{g(\bfs)}\right],
\]
so that the analog of our approximation method, Eq.~(\ref{eq:iter}) is
\[
g(\bfs) =
\frac{e^{-\beta H(\bfs)/2}}
{\sqrt{1 + \frac{1}{N} \sum_i^N\frac{\hat{\delta}_i g(\bfs)}{g(\bfs)}}}.
\]
Solving this involves $\calo (N)$ work at the very least and will yield
$N$ values of $g$, one for each proposed new state $\hat{\theta}_i\bfs$.
We can sample from these to choose a site whose spin is then flipped
with probability unity.  As with our five-state problem, this is likely
to become impossible when one state is much more probable than any of
its neighbors, and this is likely to happen for sufficiently low
temperature.  A more fundamental problem is that the method as stated
requires $\calo (N)$ work to flip one spin (albeit with unit
probability).  This is clearly prohibitive, but it may be possible to
drastically improve this situation by more clever choices of the trial
PDF.  In particular, it would be interesting to study the relationship
of this method to the various cluster algorithms that are known for the
Ising model~\cite{bib:barkema}.

In conclusion, the integral equation, Eq.~(\ref{eq:inteq}), is a very
useful way to frame the problem of developing a Markov process with
specified equilibrium and trial PDF.  The method is not a panacea, in
that it will not always be possible to find a solution for $g(\bfx)$
with definite sign.  The breakdown of the method when $g$ has indefinite
sign is reminiscent of the {\it sign problem} that is sometimes
encountered in quantum Monte Carlo simulations~\cite{bib:sign}, and it
would be interesting to study this relationship further.  Moreover,
unless a trial PDF is chosen very judiciously, the amount of work
involved in finding $g$ may mitigate the advantage of unconditional
acceptance, as with the Ising model example above.  In any case, we have
shown that such a solution may be constructed perturbatively for
sufficiently small step sizes in a continuous state space, and this
observation ought to have immediate application to molecular Monte Carlo
problems, as we have also shown.  In addition, our application of this
methodology to stochastic lattice Boltzmann methods is presented in one
of the references~\cite{bib:entlbe}.

It is a pleasure to acknowledge illuminating conversations with Frank
Alexander, Peter Coveney, Mal Kalos, Bill Klein, Claudio Rebbi, Bob Swendsen
and Jeff Yepez, and careful critical readings of the draft by Harvey Gould
and Mark Novotny.  The author was supported in part by an IPA agreement with
Phillips Laboratory, and in part by the United States Air Force Office of
Scientific Research under grant number F49620-95-1-0285.

\end{document}